\documentclass{article}

\usepackage{amssymb}
\usepackage{graphicx}% Include figure files
\usepackage{dcolumn}% Align table columns on decimal point
\usepackage{bm}% bold math
\usepackage{amsmath}

\begin{document}

\title{\bf Natural Cutoffs via Compact Symplectic Manifolds}
\author{K. Nozari$^{1}$\thanks{email: knozari@umz.ac.ir}
\,,\hspace{.2cm}M. A. Gorji$^1$\thanks{{email:
m.gorji@stu.umz.ac.ir}}\,,\hspace{.2cm}V. Hosseinzadeh$^1$
\thanks{{email: v.hosseinzadeh@stu.umz.ac.ir}}\hspace{.2cm}
and\hspace{.2cm} B. Vakili$^2$\thanks{email:
b-vakili@iauc.ac.ir}\vspace{.15cm}\\$^1${\small{\it
Department of Physics, Faculty of Basic Sciences, University
of Mazandaran,}}\\{\small{\it P. O. Box 47416-95447, Babolsar,
Iran}}\\$^2${\small {\it Department of Physics, Central
Tehran Branch, Islamic Azad University, Tehran, Iran.}}}

\maketitle
\begin{abstract}
In the context of phenomenological models of quantum gravity, it is claimed
that the ultraviolet and infrared natural cutoffs can be realized from
local deformations of the Hamiltonian systems. In this paper, we scrutinize
this hypothesis and formulate a cutoff-regularized Hamiltonian system. The
results show that while local deformations are necessary to have cutoffs,
they are not sufficient. In fact, the cutoffs can be realized from
globally-deformed Hamiltonian systems that are defined on compact
symplectic manifolds. By taking the universality of quantum gravity
effects into account, we then conclude that quantum gravity cutoffs are
global (topological) properties of the symplectic manifolds. We justify our
results by considering three well-known examples: The Moyal, Snyder and
polymer deformed Hamiltonian systems.
\vspace{5mm}
\end{abstract}

\section{Introduction}
General relativity describes our Universe with high accuracy, but it fails to be
applicable in the very high and low energy regimes. In the standard Big Bang cosmology,
the initial value problem appears when one utilizes general relativity to describe
the early state of the Universe \cite{IVP}. Also, cosmological observations indicate
that Universe is in an accelerating phase with an unknown origin usually dubbed as the
dark energy \cite{DarkEnergy}. The initial value problem can be alleviated by
considering an inflationary phase before nucleosynthesis era \cite{Inflation}. But it
is an artificial scenario that is glued to the standard model of cosmology. Indeed,
both of the initial value and dark energy problems appear in the high and low energy
regimes respectively when quantum effects are important as well as gravitational
effects. It is then natural to expect that these problems will be addressed in the
context of ultimate quantum theory of gravity. In the absence of such a conclusive
theory, quantum gravity candidates such as string theory and loop quantum gravity
suggest the existence of a minimal length scale (and also maximal energy scale)
leading to a universal ultraviolet (UV) cutoff \cite{String,Loop}. Interestingly, Big
Bang singularities are replaced by quantum bounces in quantum cosmology scenarios
\cite{LQC}. Furthermore, in the context of black hole physics, holographic principle
is suggested for quantum gravity proposal which can naturally explain the late time
cosmic acceleration through the existence of an infrared (IR) cutoff \cite{Holographic}.
Existence of these natural cutoffs is essential for the regularization of quantum field
theories \cite{GUP-UVIR}. Evidently, taking a minimal length scale into account
naturally makes quantum field theories to be UV-regularized
\cite{Snyder-QFT,NCGFT,NCGFT2,Kempf,GUP-UV}. Also, it is shown that the existence of
an IR cutoff is necessary for the quantization of fields in curved spacetimes
\cite{Brandenberger}. Thus, modification of general relativity in the high and low
energy regimes seems to be necessary in order to take into account these natural
quantum gravity cutoffs.

At the flat limit (when gravity is negligible) of a final quantum gravity theory, one
thus expects to find a non-gravitational theory which includes natural UV and IR
cutoffs. The special relativity, however, does not support the existence of these
cutoffs. The doubly special relativity theories are then proposed as candidates for
the flat limit of quantum gravity which take into account a minimal
observer-independent length scale, as a UV cutoff, in special relativity
\cite{DSR,DSR2}. The triply special relativity is also formulated to include an IR
cutoff in doubly special relativity theories \cite{TSR}. The doubly special relativity
theories are formulated on curved momentum spaces with de Sitter and anti-de Sitter
geometries \cite{dS,AdS} and the duality of curved momentum spaces with noncommutative
spacetimes is also shown in the context of quantum geometry \cite{Majid}. Indeed,
noncommutative geometry is an appropriate framework to formulate theories which deal
with UV and IR cutoffs \cite{NCQM} (see Refs. \cite{NCGFT2,Seiberg,Susskind,NC-UVIR}
for the UV/IR mixing effect). The other effective phenomenological models of quantum
gravity such as the generalized uncertainty principle that is inspired by string
theory \cite{Kempf,GUP-UV,GUP} and polymer quantum mechanics \cite{Ashtekar,Corichi1}
which is investigated in the symmetric sector of loop quantum gravity are in close
relation with noncommutative spaces (see for instance \cite{poly-nc,snyder-poly}). At
the classical level, all of these models, and any model which includes natural UV and
IR cutoffs, can be realized from the deformed Hamiltonian system. Such systems are
usually led to deformed noncanonical Poisson algebras with non-vanishing commutation
relations between positions and momenta which signal the existence of UV and IR
cutoffs respectively \cite{NC-class}. This is, however, a local criterion and we know
that the Hamiltonian system are described by the symplectic manifolds which are locally
equivalent. Any noncanonical Poisson algebra then can be transformed to a canonical
form in the light of the Darboux theorem. The question then arises: What is the origin
of cutoff? In this paper we consider the Hamiltonian systems in their most fundamental
form in order to find what is the origin of cutoffs and when a Hamiltonian system is
cutoff-regularized.

\section{Hamiltonian Systems}
In order to explore which properties a Hamiltonian system should have to be
cutoff-regularized, we briefly review the general formalism of Hamiltonian systems in
the language of symplectic geometry in this section.

Suppose that ${\mathcal{Q}}$ is a configuration space manifold of a mechanical system,
the Hamiltonian system is defined on cotangent bundle $T^{\ast}{\mathcal{Q}}$ which is
the phase space of the system under consideration. Cotangent bundle admits symplectic
structure and, therefore, the phase space is naturally a symplectic manifold. The
symplectic manifold $({\mathcal M},\Omega)$ is a manifold ${\mathcal M}$ with symplectic
structure $\Omega$ which is a closed non-degenerate $2$-form on ${\mathcal{M}}$. Since
the symplectic structure is non-degenerate, one can assign a vector field to a function
$f:{\mathcal M}\rightarrow\,{\mathbb R}$ as $\Omega(X_{_f})=df$. The Poisson bracket
between two real-valued functions $f$ and $g$ is then defined as
\begin{equation}\label{PBD}
\{f,\,g\}=\Omega(X_{_f},X_{_g})\,,
\end{equation}
The closure of the symplectic structure $d\Omega=0$ ensures that the Jacobi identity is
satisfied by the resultant Poisson bracket. The symplectic structure or equivalently the
Poisson bracket (\ref{PBD}) properly define kinematics on the phase space.

The Hamiltonian system is defined by the triple $({\mathcal{M}},\Omega,X_{_H})$, where
$X_{_H}$ is the Hamiltonian vector field which solves the dynamical equation
\begin{equation}\label{HVF}
\Omega(X_{_H})=dH\,,
\end{equation}
where $H:{\mathcal{M}}\rightarrow\,{\mathbb{R}}$ is the Hamiltonian function. The
integral curves of the Hamiltonian $X_{_H}$ are the solutions of the Hamilton's
equations which determine the evolution of the system.

The symplectic manifolds are even-dimensional and are oriented by the Liouville volume.
The Liouville volume for $2n$-dimensional symplectic manifold ${\mathcal{M}}$ is defined
as
\begin{eqnarray}\label{Liouville-Volume}
\Omega^n=\frac{(-1)^{n(n-1)/2}}{n!}\,\Omega\wedge...\wedge\Omega
\hspace{1cm}(n\,\,\mbox{times}),
\end{eqnarray}
and the Liouville theorem states that the volume (\ref{Liouville-Volume}) is conserved
along the Hamiltonian flow of $X_{_H}$ as
\begin{equation}\label{Liouville-Theorem}
{\mathcal{L}}_{X_{_H}}\Omega^n=0\,,
\end{equation}
where ${\mathcal{L}}_{X_{_H}}\Omega^n$ denotes the Lie derivative of $\Omega^n$ along
$X_{_H}$. The relation (\ref{Liouville-Theorem}) can be easily deduced from the fact
that the Lie derivative of the symplectic structure along the Hamiltonian vanishes:
${\mathcal{L}}_{X_{_H}}\Omega=X_{_H}(d\Omega)+d(X_{_H}(\Omega))=0$, where we have used
(\ref{HVF}) and the fact that $d\Omega=0$.
\subsection{Standard Hamiltonian systems}
The standard Hamiltonian systems are special case of the above definition. More
precisely, the standard Hamiltonian system $({\mathcal M}_0,\Omega_0,X_{_{H_0}})$ is
defined on trivial topology ${\mathcal M}_0={\mathbb R}^{2n}$ that lead to two the
features: i) The closed 2-form $\Omega_0$ is also exact, ii) The manifold
${\mathcal M}_0$ can be completely covered by one chart\footnote{There are also
Hamiltonian systems that are defined on nontrivial topology such as the system of
magnetic dipoles with compact $S^2$ phase space. Note, however, that this phase space is
not cotangent bundle and it is indeed a classical limit of a quantum system rather than
a purely classical system. In this paper, we deal with purely classical systems where
the associated Hamiltonian systems are defined on cotangent bundle of a configuration
space.}. There are many charts over ${\mathcal M}_0$, but, of more importance is the
chart $U_0\subset{\mathcal M}_0$ in which the local coordinates are the positions and
momenta of particle. In this chart, the symplectic structure takes the canonical form
\begin{equation}\label{omega0}
\Omega_0|_{_{U_0}}=dq_0^i\wedge d{p_0}_i\,,
\end{equation}
where the coordinates $(q_0^i,{p_0}_i)$ are the positions and momenta of the particles
respectively and $i=1,2,..,n$. The variables $(q_0^i,{p_0}_i)$ are known as the
canonical variables which for the standard non-deformed Hamiltonian systems coincide
with the positions and momenta of particles. But this coincidence may fail in general
(for instance in the context of phenomenological quantum gravity models which are the
subject of the next subsection). The standard well-known local form of the Poisson
bracket can be deduced by using (\ref{omega0}) into the definition (\ref{PBD}) which
leads to the canonical Poisson algebra
\begin{eqnarray}\label{PA0}
\{q_0^i,q_0^j\}=0,\hspace{1.5cm}\{q_0^i,{p_0}_j\}=\delta^i_j,\hspace{1.5cm}
\{{p_0}_i,{p_0}_j\}=0.
\end{eqnarray}
Solving equation $\Omega_0(X_{_{H_0}})=dH_0$ for the Hamiltonian vector field, gives
the following solution
\begin{eqnarray}\label{HVF0}
X_{_{H_0}}|_{_{U_0}}=\frac{\partial{H_0}}{\partial{p_0}_i}\frac{\partial}{
\partial{q_0^i}}-\frac{\partial{H_0}}{\partial{q_0^i}}\frac{\partial}{
\partial{p_0}_i}.
\end{eqnarray}
The integral curves of the Hamiltonian vector field are the solutions of the Hamilton's
equations which are given by
\begin{eqnarray}\label{Hamilton0}
\frac{dq_0^i}{dt}=\frac{\partial{H_0}}{\partial{p_0}_i},\hspace{1.5cm}
\frac{d{p_0}_i}{dt}=-\frac{\partial{H_0}}{\partial{q_0^i}}.
\end{eqnarray}
The symplectic structure takes always the canonical form (\ref{omega0}) in terms of
positions and momenta of particles for the standard Hamiltonian system
$({\mathcal M}_0,\Omega_0,X_{_{H_0}})$ and the different physical systems are
classified by the different Hamiltonian functions (or equivalently the different
forces that are exerted to the systems). Therefore, the trajectories on ${\mathcal M}_0
$, which are determined by the Hamilton's equations (\ref{Hamilton0}), are different
through the different Hamiltonian functions. In other words, the standard physical
systems described by the Hamiltonian system $({\mathcal M}_0,\Omega_0,X_{_{H_0}})$ are
globally the same and locally different.

Here, the Hamiltonian system $({\mathcal M}_0,\Omega_0,X_{_{H_0}})$ is represented in
physical chart $U_0\subset{\mathcal M}_0$ in which the local coordinates are the
positions and momenta of particles in terms of which the Poisson algebra and Hamiltonian
vector field takes the local forms (\ref{PA0}) and (\ref{HVF0}) respectively. However,
one could also work in another chart $U_0'$ with local coordinates $(u^i,v_i)$ in terms
of which the Poisson algebra becomes noncanonical and the Hamiltonian vector field and
Hamiltonian function also take different functional forms. The well-known example of
such local transformation is a system affected by a magnetic field. In this system it
is better (in some sense) to work with noncanonical variables $(u^i,v_i)$ rather than
the standard canonical ones $(q_0^i,{p_0}_i)$ (see chapter six of Ref. \cite{Marsden}
for more details). All the physical results (such as the expectation value of a
physical observable) are the same in two charts $U_0$ and $U_0'$.

\subsection{Deformed Hamiltonian systems}
In this subsection, we explore a cutoff-regularized Hamiltonian system to be quantum
gravity counterpart of the standard Hamiltonian system defined in the pervious
subsection. In other words, we would like to formulate a fundamental Hamiltonian system
that is cutoff-regularized and also reduces to the standard Hamiltonian system at the
low energy regime (correspondence principle).
\subsubsection{Noncanonical representation}
We consider general Hamiltonian system $({\mathcal M},\Omega,X_{_H})$ as a quantum
gravity counterpart of the standard Hamiltonian system
$({\mathcal M}_0,\Omega_0,X_{_{H_0}})$. These two Hamiltonian systems may be different
from both of the local and global points of view. In order to preserve the
correspondence principle, we start from local deformation of Poisson algebra that also
is the starting point of phenomenological quantum gravity model. In these models, the
Poisson algebra is usually noncanonical in terms of positions and momenta of particles.
Thus, in physical chart $U\subset{\mathcal M}$ in which the local coordinates are the
positions $q^i$ and momenta $p_i$ of the particles\footnote{We use the the notation
$U$ for physical chart on ${\mathcal M}$ while we use the notation $U_0$ for physical
chart on ${\mathcal M}_0$. Note that these two charts are defined on different
manifolds, but they are same from the physical point of view since in both of them the
local coordinates are particles' positions and momenta ($(q^i,p_i)$ for $U$ on
${\mathcal M}$ and $(q_0^i,{p_0}_i)$ for $U_0$ on ${\mathcal M}_0$). This is the reason
for which we call both of them as physical chart. At the low energy regime, where the
two Hamiltonian systems coincide, the set of variables $(q^i,p_i)$ and $(q_0^i,{p_0}_i)
$ are also coincide. Only in this regime, $U$ and $U_0$ are completely the same charts
from both of the physical and mathematical points of view.}, we suppose that the
symplectic structure takes the following general noncanonical form
\begin{eqnarray}\label{omega}
\Omega|_{_U}=dq^i\wedge dp_i-\epsilon\Big(\sigma_i^j dq^i\wedge dp_j+\frac{1}{2}
\alpha_{ij}dq^i\wedge dq^j+\frac{1}{2}\beta^{ij}dp_i\wedge dp_j\Big),
\end{eqnarray}
where $\sigma_j^i(q,p)$, $\alpha_{ij}(q,p)$ and $\beta^{ij}(q,p)$ are arbitrary functions
of $(q^i,p_i)$ and $i,j=1,...,n$. The condition $d\Omega=0$ gives however the following
constraints on them
\begin{eqnarray}\label{Jacobi}
\frac{\partial\sigma^i_j}{\partial q^{k}}\,-\,\frac{\partial\sigma^i_k}{
\partial q^j}\,+\,\frac{\partial\alpha_{kj}}{\partial p_i}=0,\hspace{2cm}
\frac{\partial\sigma^i_k}{\partial p_j}\,-\,\frac{\partial\sigma^j_k}{
\partial p_i}\,+\,\frac{\partial\beta^{ij}}{\partial q^k}=0,\\
\frac{\partial\alpha_{ij}}{\partial q^k}+\frac{\partial \alpha_{jk}}{
\partial q^i}+\frac{\partial\alpha_{ki}}{\partial q^j}=0,\hspace{2cm}
\frac{\partial\beta^{ij}}{\partial p_k}+\frac{\partial\beta^{jk}}{
\partial p_i}+\frac{\partial\beta^{ki}}{\partial p_j}=0\,,\nonumber
\end{eqnarray}
which should be held to guaranty that the Jacobi identity is satisfied by the associated
Poisson Bracket. While the closure of $\Omega$ is guarantied by the above relations, it
may be not exact. This is because that the exactness will be determined by the
integration of functions $\sigma_j^i(q,p)$, $\alpha_{ij}(q,p)$ and $\beta^{ij}(q,p)$ on
${\mathcal M}$ that may have nontrivial topology. Note that $\Omega_0$ is a closed and
also exact 2-form on ${\mathcal M}_0={\mathbb R}^{2n}$ for the standard Hamiltonian
system $({\mathcal M}_0,\Omega_0,X_{_{H_0}})$.

In order to find the Poisson bracket related to the proposed symplectic structure
(\ref{omega}), it is useful to represent $\Omega$ as a matrix {\boldmath$\Omega$}. The
components are defined as $ $\mbox{\boldmath$\Omega$}$_{_{IJ}}=\Omega(\frac{\partial}{
\partial z^I},\frac{\partial}{\partial z^J})$ while $z^I=q^I$ for $I=1,...,n$ and $z^I=
p^I$ for $I=n+1,...,2n$. Then we have
\begin{eqnarray}\label{matrix-omega}
\mbox{\boldmath$\Omega$}|_{_U}\doteq
\begin{pmatrix}
\epsilon $\mbox{\boldmath${\alpha}$}$&$\mbox{\boldmath$1$}$-\epsilon$\mbox{\boldmath${\sigma}$}$\\
-$\mbox{\boldmath$1$}$+\epsilon$\mbox{\boldmath${\sigma}$}$&\epsilon$\mbox{\boldmath${\beta}$}$
\end{pmatrix}\,,
\end{eqnarray}
where $\mbox{\boldmath${\alpha}$}_{ij}=\alpha_{ij}$, $\mbox{\boldmath${\sigma}$}_{ij}=
\sigma_i^j$,$\mbox{\boldmath${\beta}$}_{ij}=\beta^{ij}$ and $\mbox{\boldmath${1}$}_i^j=
\delta_i^j$ are $n\times n$ matrices. The Poisson brackets worked out as
\begin{equation}\label{PBD2}
\{f,\,g\}|_{_U}=f(z)\,\overleftarrow{\partial}_I\mbox{\boldmath${\mathcal P}$}^{IJ}
\overrightarrow{\partial}_J\,g(z)\,,
\end{equation}
where \mbox{\boldmath${\mathcal P}$} is the matrix representation of the Poisson tensor
which satisfies matrix relation $\mbox{\boldmath${\mathcal P}$}\mbox{\boldmath$\Omega$}=
-$\mbox{\boldmath${1}$}\footnote{The phenomenological models of quantum gravity usually
start from definition (\ref{PBD2}) and then locally deforms the Poisson tensor
which leads to a locally-deformed Poisson algebra. For our purpose, we could consider a
Poisson manifold rather than a symplectic one and start from the relation (\ref{PBD2})
and deforms it in such a way that it leads to a general locally-deformed Poisson algebra
(see Ref. \cite{NC-class} for more details).}. Thus it is actually the inverse of
$\mbox{\boldmath$\Omega$}$ up to a sign and the problem of finding the Poisson bracket
boils down to the problem of finding \mbox{\boldmath${\mathcal P}$}. Considering the
matrix representation of $\Omega$ as $\mbox{\boldmath${\Omega}$}=
$\mbox{\boldmath${\Omega_0}$}$-\epsilon$\mbox{\boldmath${\Omega'}$}$ $ and using the
fact that $\mbox{\boldmath${\Omega^{-1}}$}=$\mbox{\boldmath${\Omega_0^{-1}}$}$+\epsilon
\,$\mbox{\boldmath${\Omega_0^{-1}}$}$ $\mbox{\boldmath${\Omega'}$}$
$\mbox{\boldmath${\Omega_0^{-1}}$}$+{{\mathcal O}(\epsilon^2)}$, we find
\mbox{\boldmath${\mathcal P}$} up to first order of $\epsilon$ as
\begin{eqnarray}\label{matrix-p}
\mbox{\boldmath${\mathcal P}$}|_{_U}\doteq
\begin{pmatrix}
\epsilon$\mbox{\boldmath${\beta}$}$&$\mbox{\boldmath$1$}$+\epsilon$\mbox{\boldmath${\sigma}$}$\\
-($\mbox{\boldmath$1$}$+\epsilon$\mbox{\boldmath${\sigma}$}$)&\epsilon$\mbox{\boldmath${\alpha}$}$
\end{pmatrix}\,.
\end{eqnarray}
From the relations (\ref{PBD2}) and (\ref{matrix-p}), the Poisson bracket is given by
\begin{eqnarray}\label{NCPB}
\{f,\,g\}|_{_U}=\big(\delta_j^i+\epsilon\,\sigma_j^i\big)\Big(\frac{\partial f}{
\partial q^i}\frac{\partial g}{\partial p_j}-\frac{\partial f}{\partial p_j}\frac{
\partial g}{\partial q^i}\Big)+\epsilon\,\Big(\beta^{ij}\frac{\partial f}{\partial q^i}
\frac{\partial g}{\partial q^j}+\alpha_{ij}\frac{\partial f}{\partial p_i}\frac{
\partial g}{\partial p_j}\Big).
\end{eqnarray}
Note that the Poisson bracket is globally defined in (\ref{PBD}) and it is represented
on local physical chart $U\subset{\mathcal M}$ in the relation (\ref{PBD2}). From the
relation (\ref{NCPB}), the locally-deformed noncanonical Poisson algebra associated to
the symplectic structure (\ref{omega}) in chart $U$ will be
\begin{eqnarray}\label{NPA}
\{q^i,q^j\}=\epsilon\,\beta^{ij}(q,p),\hspace{1.5cm}{\{q^i,p_j\}=\delta^i_j+\epsilon\,
\sigma^{i}_{j}(q,p)},\hspace{1.5cm}\{p_i,p_j\}=\epsilon\,\alpha_{ij}(q,p).
\end{eqnarray}
The above locally-deformed Poisson algebra is the starting point of almost all of the
phenomenological models of quantum gravity\footnote{There is also another interesting
approach to take into account UV and IR cutoffs that is the so-called Moyal star product.
This approach is fundamentally different from the setup we presented in this paper. The
deformed algebra (\ref{NPA}) can be also deduced in Moyal picture by considering the
noncommutative product law between two arbitrary functions $f,g:{\mathcal M}\rightarrow
\,{\mathbb R}$ as \cite{NCQM}
\begin{eqnarray}\label{Moyal-product}
f(z)\ast g(z)=f(z)\exp\Big(\frac{1}{2}{{\overleftarrow{\partial}_I}\,
{\mbox{\boldmath${\mathcal P}$}^{IJ}}\,{\overrightarrow{\partial}_J}}\Big)g(z),
\end{eqnarray}
where $\mbox{\boldmath${\mathcal P}$}^{IJ}$ are the components of the Poisson matrix
(\ref{matrix-p}). Defining Moyal bracket as $[f,g]=f\ast{g}-g\ast f$, one can easily
obtain the following deformed noncommutative Moyal algebra
\begin{eqnarray}\label{NMA}
[q^i,q^j]=\epsilon\,\beta^{ij}(q,p),\hspace{1.5cm}{[q^i,p_j]=\delta^i_j+\epsilon\,
\sigma^{i}_{j}(q,p)},\hspace{1.5cm}[p_i,p_j]=\epsilon\,\alpha_{ij}(q,p).
\end{eqnarray}
It is important to note that while the two algebras (\ref{NPA}) and (\ref{NMA}) seem to
be the same, they are fundamentally different. In fact, the algebra of the functions is
always commutative in our setup, while it is noncommutative in the spirit of the Moyal
product law (\ref{Moyal-product}). These two different pictures are related to each
other through the Seiberg-Witten map.}. Indeed, the non-vanishing commutation relations
between coordinates $q^i$ signal the existence of a UV cutoff which is here labeled
by $\beta^{ij}$ functions. This fact is first suggested by Heisenberg itself in 1938 in
order to remove the UV divergences in quantum field theory \cite{Heisenberg}. In 1947,
Snyder formulated a discrete Lorentz-invariant spacetime \cite{Snyder} and it was shown
that quantum field theories are UV-regularized in this setup \cite{Snyder-QFT}. In the
same manner, one can consider non-vanishing commutation relations between momenta $p_i$,
which is determined by $\alpha_{ij}$ in our setup, in order to include an IR cutoff.
Existence of such an IR cutoff is essential for the renormalization of quantum fields
in curved spacetime \cite{Brandenberger}. The locally-deformed Poisson algebra
(\ref{NPA}) could then make the system under consideration to be UV/IR-regularized.
This idea recently suggested by many phenomenological models of quantum gravity.
Inspired by string theories, the generalized uncertainty principle was suggested which
can be realized from the deformed Poisson algebra, as a special case of (\ref{NPA}),
in the semiclassical regime \cite{GUP-classical}. The UV/IR-regularized Snyder-deformed
Poisson algebra is suggested which can be also considered as a special case of
(\ref{NPA}) \cite{Snyderclass}. The doubly special relativity theories was also
suggested which take into account a minimal observer-independent length scale in special
relativity as an UV cutoff \cite{DSR}. Evidently the commutation relation between the
positions becomes non-vanishing for the corresponding deformed Poincar\'{e} algebra
\cite{DSR2,Glikman}. Considering the non-vanishing Poisson brackets between the
four-momenta leads to the triply special relativity which take into account an IR
cutoff in doubly special relativity theories\footnote{It should be noted that the
doubly special relativity theories are formulated on the extended phase space in a
relativistic manner \cite{DSR-RL} while our setup is formulated on the reduced phase
space. The main idea of the paper is, however, applicable for these theories and it is
possible to generalize this setup to case of extended phase space (which is the
cotangent bundle of spacetime) and also the more interesting case of curved spacetime
\cite{DSR-HS}.} \cite{TSR}.

In order to obtain the deformed Hamilton's equations in this setup, substituting
(\ref{omega}) into the relation (\ref{HVF}) gives
\begin{eqnarray}\label{A2}
\left\{
\begin{array}{ll}
-y_i+\epsilon\big(\alpha_{ij}\,x^j+\sigma^j_i\,y_j\big)=\frac{\partial{H}}{\partial q^i},\\\\
x^i+\epsilon\big(-\sigma^i_j\,x^j+\beta^{ij}\,y_j\big)=\frac{\partial{H}}{\partial p_i},
\end{array}
\right.
\end{eqnarray}
for a vector field $X_{_H}|_{_U}=x^i\frac{\partial}{\partial{q^i}}+y_i\frac{\partial
}{\partial{p_i}}$. Since the structure (\ref{omega}) is non-degenerate, the coupled
set of equations (\ref{A2}) has a unique solution. While finding an exact solution
of the above coupled equations is not an easy task, the approximated solution is
indeed sufficient for our purpose. To first order of $\epsilon$, the solution is
\begin{eqnarray}\label{NCVF}
X_{_H}|_{_U}=\bigg[\frac{\partial H}{\partial p_i}+\epsilon\Big(\sigma^i_j\frac{
\partial H}{\partial p_j}+\beta^{ij}\frac{\partial H}{\partial q^j}\Big)
\bigg]\frac{\partial}{\partial q^i}-\bigg[\frac{\partial H}{\partial q^i}-
\epsilon\Big(-\sigma^j_i\frac{\partial H}{\partial q^j}+\alpha_{ij}\frac{
\partial H}{\partial p_j}\Big)\bigg]\frac{\partial}{\partial p_i}.
\end{eqnarray}
The integral curves of the Hamiltonian vector field (\ref{NCVF}) are the solutions of the
Hamilton's equations which are given by
\begin{eqnarray}\label{Hamilton}
\frac{dq^i}{dt}=\frac{\partial H}{\partial p_i}+\epsilon\Big(\sigma^i_j
\frac{\partial H}{\partial p_j}+\beta^{ij}\frac{\partial{H}}{\partial{
q^j}}\Big),\nonumber\\{\quad}\frac{dp_i}{dt}=-\frac{\partial{H}}{
\partial{q^i}}+\epsilon\Big(-\sigma^j_i\,\frac{\partial{H}}{\partial{
q^j}}+\alpha_{ij}\frac{\partial{H}}{\partial{p_j}}\Big).
\end{eqnarray}
The Hamiltonian system $({\mathcal M},\Omega,X_{_H})$ is represented in physical chart
$U\in{\mathcal M}$ in this subsection. All of the above relations reduce to the standard
ones in the limit of $\epsilon\rightarrow\,0$. In this limit, the two charts $U$ and
$U_0$ are the same that are defined on the same manifold ${\mathcal M}\rightarrow
{\mathcal M}_0$ with same coordinates $(q^i,p_i)\rightarrow(q_0^i,{p_0}_i)$ and the
relations (\ref{omega}), (\ref{NPA}), (\ref{NCVF}), and (\ref{Hamilton}) reduce exactly
to their non-deformed counterparts (\ref{omega0}), (\ref{PA0}), (\ref{HVF0}), and
(\ref{Hamilton0}) respectively. Thus, the defined Hamiltonian system respects the
correspondence principle.
\subsubsection{Canonical (Darboux) representation}
According to the Darboux theorem, for each point $p\in{\mathcal M}$ there is a local
chart about $p$ in which the $2$-form structure is constant. Therefore, it
is always possible to find a chart in which any symplectic structure such as the
locally-deformed symplectic structure (\ref{omega}) takes the canonical form
\cite{Marsden,Arnold}. In this respect, there exists a Darboux transformation
\begin{equation}\label{Darb-trans}
(q^i,p_i)\rightarrow\,(X^i,Y_i)
\end{equation}
from chart $U$ with noncanonical coordinates $(q^i,p_i)$ to the chart $U'$ with
canonical variables $\left(X^i(q,p),Y_i(q,p)\right)$ in which the $2$-form structure
(\ref{omega}) takes the canonical form
\begin{equation}\label{StrucDarb}
\Omega|_{_{U'}}=dX^i\wedge dY_i.
\end{equation}
Using the above local representation of symplectic manifold, from the definition
(\ref{PBD}), it is easy to show that the canonical variables $(X,Y)$ obey canonical
Poisson algebra
\begin{eqnarray}\label{Dar-CPA}
\{X^i,X^j\}=0,\hspace{1.5cm}\{X^i,Y_j\}=\delta^i_j,\hspace{1.5cm}
\{Y_i,Y_j\}=0.
\end{eqnarray}
Both of the above Poisson algebra and the standard one (\ref{PA0}) are canonical. But
it is important to note that the standard Poisson algebra is canonical in terms of
positions and momenta of particles while the algebra (\ref{Dar-CPA}) is canonical in
terms of some phase space variables. The Poisson algebra (\ref{Dar-CPA}) is the local
representation of the Hamiltonian system $({\mathcal M},\Omega,X_{_H})$ which took
noncanonical form (\ref{NPA}) in terms of the positions and momenta of particles.

Substituting (\ref{StrucDarb}) into (\ref{HVF}), gives the following local form for
the Hamiltonian vector field in this chart
\begin{eqnarray}\label{Dar-VF}
X_{_H}|_{_{U'}}=\frac{\partial H}{\partial Y_i}\frac{\partial}{\partial
X^i}-\frac{\partial H}{\partial X^i}\frac{\partial}{\partial Y_i}.
\end{eqnarray}
The integral curves of (\ref{Dar-VF}) are the solutions of the Hamilton's equations in
this chart which are given by
\begin{eqnarray}\label{Dar-Hamilton}
\frac{dX^i}{dt}=\frac{\partial{H}}{\partial Y_i},\hspace{2cm}
\frac{dY_i}{dt}=-\frac{\partial{H}}{\partial X^i}.
\end{eqnarray}
Clearly the functional form of the Hamiltonian function should be changed as $H|_{_{U'}}
=H(\epsilon;X,Y)$ to ensure that the trajectories in two coordinates charts $(X^i,Y_i)
\in{U'}$ and $(q^i,p_i)\in{U}$ coincide on ${\mathcal M}$ which are determined by the
relations (\ref{Hamilton}) and (\ref{Dar-Hamilton}). This is because the dynamical
equation (\ref{HVF}) should be satisfied in chart-independent manner on ${\mathcal M}$.
In other words, we have a unique Hamiltonian system $({\mathcal M},\Omega,X_{_H})$
which is represented in two different charts $U$ and $U'$. It is important to note that
$(q^i,p_i)\in{U}$ are the positions and momenta of particles by definition while the
canonical coordinates $(X^i,Y_i)$ are just some variables on chart $U'$ on
${\mathcal M}$. Nevertheless, there is not any thing to stop someone to work with
$(X^i,Y_i)$ rather than the positions and momenta $(q^i,p_i)$\footnote{We deal with
classical systems. But, it is important to note that the quantization of a symplectic
structure is not unique. See for instance Refs. \cite{DSR-Quant} and \cite{NC-Quant}
that show this feature in the contexts of doubly special relativity and noncommutative
quantum mechanics respectively.}. Both of these variables, however, coincide with
positions $q_0^i$ and momenta ${p_0}_i$ of particles in standard Hamiltonian system
$({\mathcal M}_0,\Omega_0,X_{_{H_0}})$ in the limit of $\epsilon\rightarrow\,0$ when
the transformation (\ref{Darb-trans}) coincides with the identity map.

\section{Cutoffs and Topology}
In the pervious section we considered the Hamiltonian system
$({\mathcal M},\Omega,X_{_H})$ as the quantum gravity counterpart of the standard
Hamiltonian system $({\mathcal M}_0,\Omega_0,X_{_{H_0}})$ such that the latter being the
low energy limit $\epsilon\rightarrow0$ of the former. These two Hamiltonian systems may
be different from local and also global points of view. From local point of view, the
difference is clear since the trajectories (\ref{Hamilton0}) and (\ref{Hamilton}) are
different even if the two Hamiltonian systems being globally the same and defined on the
same manifolds ${\mathcal M}={\mathcal M}_0$. In other words,
there is not a transformation that locally transforms
these two systems to each other. From global point of view, note that while topology
of the symplectic manifold ${\mathcal M}_0$ is fixed to be ${\mathbb R}^{2n}$, topology
of ${\mathcal M}$ is undefined and may be nontrivial. We are interested in Hamiltonian
systems which take into account natural quantum gravity cutoffs. A cutoff usually
considered to be a maximal or minimal value of a phase space variable (for instance a
maximal value for the momentum corresponding to a UV cutoff). But, this is not a
precise definition from the mathematical point of view since it is a chart-dependent
criterion and one could work with another chart in which the associated local
coordinates become unbounded (for instance the inverse of the transformations that
usually implemented in general relativity on a spacetime manifold to obtain the
extended form). The cutoff then will be defined in a more precise manner. We
note that the number of microstates are determined by the volume of the phase space
and the existence of cutoffs makes the number of microstates to be finite in quantum
gravity regime. Therefore, we claim that the system is cutoff-regularized when the phase
space volume is finite\footnote{Note also that in the presence of UV cutoff, the volume of
the momentum space of the corresponding phase space becomes finite while the volume of the
position space gets finite value when an IR cutoff exists. In the presence of both of
these cutoffs (may be with UV/IR mixing), the total phase space will be finite.}. This
criterion is completely sensible and appropriate since the integration of phase space
volume over whole the phase space is an invariant (chart-independent) quantity. To have
a finite value for the phase space volume, there will be at least one chart (the Darboux
chart as we will show in this section) in which the phase space variables become bounded.
The total volume of the symplectic manifold ${\mathcal M}$ is the integration of the
Liouville volume (\ref{Liouville-Volume}) over whole the manifold as
\begin{equation}\label{Tot-Vol}
\mbox{Vol}(\Omega^n)=\int_{_{\mathcal M}}\,\Omega^n\,.
\end{equation}
This is a global object over the manifold and one could calculate it in an arbitrary
chart over ${\mathcal M}$. We focus on (\ref{Tot-Vol}) in two charts $U$ and $U'$ which
we considered in the pervious section and explore how total phase space volume
(\ref{Tot-Vol}) could be finite\footnote{To be more precise, one should consider covering
$\{U_i\}$ of ${\mathcal M}$ that covers each point of manifold and then find a partition
of unity subordinate to $\{U_i\}$ to calculate (\ref{Tot-Vol}). Here, however, we deal
with maximally symmetric manifolds such as $S^n$ which are completely covered with two
charts. For these particular cases, one could also work with one chart but be careful
that it is singular at one point.}. In local chart $U$ with noncanonical coordinates
$(q,p)$ (which are the positions and momenta of particles by definition) with the local
form (\ref{omega}) of the symplectic structure, (\ref{Tot-Vol}) takes the local form
\begin{equation}\label{TVol-NCdet}
\mbox{Vol}(\Omega^n)|_{_U}=\int_{_{\mathcal M}}
\sqrt{\det\mbox{\boldmath$\Omega$}(q,p)}\,d^nq\wedge d^np\,,
\end{equation}
where $\mbox{\boldmath$\Omega$}$ is given by the relation (\ref{matrix-omega}). The
non-degeneracy of the symplectic structure implies $\det$\mbox{\boldmath$\Omega$}$
\neq0$ and the explicit form of $\det\mbox{\boldmath$\Omega$}|_{_U}=\det
$\mbox{\boldmath$\Omega$}$(q,p)$ should be determined just after specifying special
forms to the functions $\sigma_i^j(q,p)$, $\alpha_{ij}(q,p)$ and $\beta^{ij}(q,p)$.
Note that $\det$\mbox{\boldmath$\Omega$}$(q,p)\neq1$ and also it is explicitly
function of coordinates $(q,p)$ in this chart. Choosing appropriate form for the
functions $\sigma_i^j(q,p)$, $\alpha_{ij}(q,p)$ and $\beta^{ij}(q,p)$, the total
volume (\ref{Tot-Vol}) can be finite through the local representation
(\ref{TVol-NCdet}). Indeed, the phenomenological models of quantum gravity suggest
specific forms for these functions in such a way that the associated total volume
becomes finite. This result is not surprising and, indeed, it is well known from the
phase spaces with deformed Hamiltonian systems such as the generalized uncertainty
principle setups \cite{GUP-classical}, polymerized phase spaces \cite{NPLYMR}, and the
doubly special relativity theories \cite{DSR-Th}. Here it is only rewritten in the
language of symplectic geometry\footnote{In the standard Hamiltonian systems with
symplectic structure (\ref{omega0}), $\det\,$\mbox{\boldmath$\Omega_0$}$(q,p)=1$ and
the positions $q_0^i$ and momenta ${p_0}_i$ of particles are unbounded variables.
Therefore, the total volume diverges. More precisely, the spatial part of the integral
(\ref{Tot-Vol}) is usually restricted to the spatial volume $V$ of the system under
consideration, but the momentum part is really diverging since there is no a priori
restriction on the momenta. We know that the total volume of the phase space determines
the number of microstates for the system under consideration and it is also well known
that the number of microstates is finite even in phase spaces with the standard
symplectic structure (\ref{omega0}). In the absence of any natural cutoff, how the
momentum part of the phase space volume gets finite value for the statistical systems?
Indeed, the ensemble densities (Dirac delta function and Boltzmann factor for the
microcanonical and canonical ensembles respectively) constraint the integral of the
total volume to give a finite phase space volume.}. Nevertheless, this is a
coordinate-dependent criterion and it seems that it can be disappeared by a suitable
coordinates transformation such as the Darboux transformation (\ref{Darb-trans}). Now,
consider the total volume (\ref{Tot-Vol}) in Darboux chart $U'$ that takes the local
form
\begin{equation}\label{Tot-VolDarb}
\mbox{Vol}(\Omega^n)|_{_{U'}}=\int_{_{\mathcal{M}}}\,d^nX\wedge d^nY\,,
\end{equation}
where we have used the fact that $\det\,$\mbox{\boldmath$\Omega$}$|_{_{U'}}=1$,
in this chart. Now, suppose that the existence of the nonconstant determinant
$\det\mbox{\boldmath$\Omega$}|_{_U}=\det\,$\mbox{\boldmath$\Omega$}$(q,p)$ makes
the total volume of the phase space to be finite by choosing appropriate forms for
the functions $\sigma_i^j(q,p)$, $\alpha_{ij}(q,p)$ and $\beta^{ij}(q,p)$ (this
is exactly the situation which arises in quantum gravity phenomenological model
which support the existence of natural cutoff, see for instance the example of
Snyder-deformed phase space in the next section), but how it could get finite in
Darboux chart $U'$ through the relation (\ref{Tot-VolDarb}). Note that the total
volume (\ref{Tot-Vol}) is an invariant quantity over ${\mathcal M}$ and should take
the same values in different charts $U$ and $U'$.

\begin{quote}
\begin{itemize}
  \item {\bf Proposition:} If the deformed noncanonical Poisson algebra
  (\ref{NPA}), represented in chart $U$ with unbounded local coordinates
  $(q^i,p_i)$, makes the total volume (\ref{Tot-Vol}) of the symplectic
  manifold ${\mathcal M}$ to be finite, there is a {\it nonlinear}
  Darboux map (\ref{Darb-trans}) from chart $U$ with coordinates $(q^i,p_i)$
  to chart $U'$ with {\it bounded} local coordinates $(X^i,Y_i)$ in terms of
  which the Poisson algebra takes the canonical form (\ref{StrucDarb}).\\

  {\it Proof}: Suppose that coordinates $(X^i,Y_i)\in{U'}$ to be unbounded
  variables. The total volume (\ref{Tot-VolDarb}) then diverges in this
  chart since $\det\,$\mbox{\boldmath$\Omega$}$|_{_{U'}}=1$. It, however,
  is an invariant quantity over ${\mathcal M}$ and took a finite value in
  chart $U$. Thus, coordinates $(X^i,Y_i)$ should be bounded in order to give
  a finite total volume. Let the transformation (\ref{Darb-trans}) being
  linear. The coordinates $(X^i,Y_i)$ will be linear combination of the
  unbounded coordinates $(q^i,p_i)$ and therefore cannot be bounded. The
  transformation (\ref{Darb-trans}) then is nonlinear. (Note that the
  existence of such transformation is guarantied by the Darboux theorem.)
\end{itemize}
\end{quote}

The above result clarifies how the cutoffs arise in the phenomenological models of
quantum gravity by means of the locally-deformed Poisson algebras of form (\ref{NPA}).
Indeed, the local deformation of the Poisson algebra (\ref{NPA}) is a chart-dependent
criterion and can be removed through a local transformation such as the Darboux
transformation (\ref{Darb-trans}). In the Darboux chart $U'$ with coordinates
$(X^i,Y_i)$, the cutoffs are still present but now show themselves through the finite
ranges of the canonical variables. It seems that the existence of cutoffs originates
from the other properties of Hamiltonian systems rather than the local deformation of
Poisson algebra such as (\ref{NPA}) that usually quantum gravitational models refer to.
If cutoffs are independent of local deformations on ${\mathcal M}$, the question
then arises: What is the origin of cutoffs? This question can be answered when one
notes that the symplectic manifolds with {\it compact topology} have finite total volume.
Indeed, if we demand a compact topology for the symplectic manifold, the Liouville
volume (\ref{Liouville-Volume}) naturally has a compact support and the total volume
(\ref{Tot-Vol}) turns out to be always finite. This criterion is coordinate-independent
and consequently there is no concern about changing the chart (local transformations).
Thus, if the total volume gets finite in a chart with local noncanonical coordinates,
it should be defined only on a compact symplectic manifold. In other words, one can
construct a locally-deformed noncanonical Poisson algebra which does not induce any
cutoff (see for instance the Moyal algebra in the next section) and also a deformed
Hamiltonian system with canonical Poisson algebra which induces a cutoff (see
polymerized phase space in the next section). Our results show that the former will be
represented on a symplectic manifold with standard ${\mathbb R}^{2n}$ topology while
the latter should be defined on a symplectic manifold with compact topology. Although
it is convenient to consider a locally-deformed noncanonical Poisson algebra in order
to include UV and IR cutoffs in the context of quantum gravity phenomenological models,
our results show that the compactness of the topology of the phase space is more
fundamental.
\begin{quote}
\begin{itemize}
 \item {\bf Definition:} The Hamiltonian system $({\mathcal M},\Omega,X_{_H})$ is
 {\it cutoff-regularized} if it is equipped with non-exact (but of course closed)
 2-form symplectic structure $\Omega$  that is defined on compact symplectic
 manifold ${\mathcal M}$.
\end{itemize}
\end{quote}
Note that the symplectic structure is closed and exact for the standard Hamiltonian
systems and it is defined on trivial ${\mathbb R}^{2n}$ topology. The compact
symplectic manifolds need to have a nontrivial second cohomology group to allow for
a closed non-degenerate $2$-form and one may read quantum gravity cutoffs of
${\mathcal{M}}$ from $H^2_{{\mathrm dR}}({\mathcal{M}})$.

Although cutoffs are originated from the global properties of the symplectic manifolds,
the local deformations (or equivalently the functional form of the functions
$\sigma_i^j(q,p)$, $\alpha_{ij}(q,p)$ and $\beta^{ij}(q,p)$) are very important from
physical point of view. To see this fact, we note that at the quantum level, the
corresponding Hilbert space takes nontrivial structure. For instance, the functions
will be periodic over compact manifold such as $S^n$ in order to respect the desired
symmetry. While the quantization of the presented setup is beyond of this work, it
is useful to consider the expectation value of the function $f:{\mathcal M}\rightarrow
\,{\mathbb{R}}$
\begin{equation}\label{Expectation-Obs}
\langle{f}\,\rangle=\frac{\int_{_{\mathcal{M}}}\,f
\,\,\Omega^n}{\int_{_{\mathcal{M}}}\,\Omega^n}\,.
\end{equation}
which is corresponding to the expectation value of an operator associated to $f$ on
the corresponding Hilbert space. Consider for instance the average of energy of the
quantum system which is the expectation value of the Hamiltonian operator. At the
classical level, it is corresponding to the expectation value of the Hamiltonian
function: $E=\langle{H}\,\rangle$. Although relation (\ref{Expectation-Obs}) is
defined in a chart-independent manner, the local functional forms of both of
the Hamiltonian function and symplectic structure (or equivalently the functional
form of $\sigma_i^j(q,p)$, $\alpha_{ij}(q,p)$ and $\beta^{ij}(q,p)$) are important to
calculate $E$. In this respect, not only the global properties but also local
deformations of Hamiltonian systems are important for a physical system. We will see
this feature explicitly for the cases of Snyder-deformed algebra and polymerized phase
space in the next section. Indeed, while both of the Snyder-deformed and polymerized
phase spaces are defined on compact momentum spaces with $S^1$ topology (and therefore
they induce a natural UV cutoff and are globally completely the same), the kinetic
energy of the system diverges for the Snyder case and it converges in the polymer
framework.

\section{Examples}
In order to justify our results, we consider three examples of deformed Hamiltonian
systems and analysis them from the local and global points of view.
\subsection{Moyal algebra}
The simplest case of deformation (\ref{omega}) to the symplectic structure emerges
when the functions $\sigma_i^j(q,p)$, $\alpha_{ij}(q,p)$ and $\beta^{ij}(q,p)$ being
constant. This particular case is known as the Moyal-deformed algebra that is defined
as \cite{Moyalclass}
\begin{equation}\label{moyal-algebra}
\{q^i,q^j\}=\theta^{ij},\hspace{1.5cm}\{q^i,p_j\}=\delta^i_j+
\eta^i_j,\hspace{1.5cm}\{p_i,p_j\}=\gamma_{ij},
\end{equation}
where $\theta^{ij}$, $\eta^i_j$, and $\gamma_{ij}$ (with $i,j=1,..,n$) are constant
functions with respect to the noncanonical variables $(q^i,p_i)$ on chart $U\subset
{\mathcal M}$\footnote{The Moyal algebra is usually defined in the spirit of the Moyal
product law (\ref{Moyal-product}) on a noncommutative phase space. But, as we say
before, we are here deal with locally-deformed noncanonical Poisson algebra on a
commutative phase space. These two pictures are related to each other by Seiberg-Witten
map.}. Comparing (\ref{moyal-algebra}) with (\ref{NPA}), one can easily find the
corresponding symplectic structure from (\ref{omega}) as
\begin{eqnarray}\label{M-twoform}
\Omega|_{_U}=\big(\delta_i^j-\eta_i^j\big)dq^i\wedge dp_j-\frac{1}{2}
\gamma_{ij}dq^i\wedge dq^j-\frac{1}{2}\theta^{ij}dp_i\wedge dp_j.
\end{eqnarray}
The constraints (\ref{Jacobi}) are automatically satisfied and the Jacobi identity
is held in this setup. Substituting (\ref{M-twoform}) into (\ref{HVF}) gives the
following solution for the Hamiltonian vector field
\begin{eqnarray}\label{M-VF}
X_{_H}|_{_U}=\bigg(\frac{\partial H}{\partial p_i}+\eta^i_j\frac{
\partial H}{\partial p_j}+\theta^{ij}\frac{\partial H}{\partial q^j}
\bigg)\frac{\partial}{\partial q^i}-\bigg(\frac{\partial H}{
\partial q^i}+\eta^j_i\frac{\partial H}{\partial q^j}-\gamma_{ij}
\frac{\partial H}{\partial p_j}\bigg)\frac{\partial}{\partial p_i},
\end{eqnarray}
which is a special case of the relation (\ref{NCVF}). The integral curves of the
Hamiltonian vector field (\ref{M-VF}) are the solutions of the Hamilton's equations
which are given by
\begin{eqnarray}\label{MHamilton}
\frac{dq^i}{dt}=\frac{\partial H}{\partial p_i}+\eta^i_j\frac{
\partial H}{\partial p_j}+\theta^{ij}\frac{\partial H}{\partial
q^j},\hspace{2cm}\frac{dp_i}{dt}=-\frac{\partial H}{\partial q^i}
-\eta^j_i\frac{\partial H}{\partial q^j}+\gamma_{ij}\frac{
\partial H}{\partial p_j}.
\end{eqnarray}
The functions $\theta^{ij}$, $\eta^i_j$, and $\gamma_{ij}$ do not explicitly depend on
the variables $(q^i,p_i)\in{U}$ and consequently the determinant of the symplectic
structure (\ref{M-twoform}) will be constant as $\det\mbox{\boldmath$\Omega$
}(\theta,\gamma)|_{_U}=1+\eta^i_i=1-\frac{1}{4}\theta^{ij}\gamma_{ij}$ but it is not
equal to unity. From the relation (\ref{TVol-NCdet}), the total volume turns out to be
\begin{equation}\label{M-Tot-Vol}
\mbox{Vol}(\Omega)|_{_U}=\int_{_{\mathcal M}}\,
\sqrt{1+\eta^i_i}\,d^nq\wedge{d^np}\,,
\end{equation}
which is clearly diverging in this setup much similar to the standard Hamiltonian
systems with symplectic structure (\ref{omega0}). Thus, the Moyal algebra
(\ref{moyal-algebra}) does not induce any cutoff on the corresponding phase space.
However, the associated Hamiltonian system is locally different from the standard
Hamiltonian system $({\mathcal M},\Omega_0,X_{_H})$. Indeed, the latter is the low
energy limit ($\theta^{ij} ,\eta^i_j,\gamma_{ij}\rightarrow0$) of the former and
there is not any local transformation that transforms these two systems to each other.
In order to clarify our results in the pervious section, we consider the Darboux
transformation (\ref{Darb-trans}) for the particular case of (\ref{moyal-algebra})
that is given by the following {\it linear} transformation
\begin{eqnarray}\label{Dar-MNA}
q^i=X^i-\frac{1}{2}\theta^{ij}Y_j,\hspace{2cm}\,p_i=Y_i+\frac{1
}{2}\gamma_{ij}X^j,
\end{eqnarray}
from chart $U$ with noncanonical variables $(q^i,p_i)$ to Darboux chart $U'$ with
canonical variables $(X^i,Y_i)$. It is easy to show that if the new variables $(X,Y)$
satisfy the standard Poisson algebra (\ref{Dar-CPA}), the variables $(q^i,p_i)$
satisfy the noncanonical Moyal algebra (\ref{moyal-algebra}) with $\eta^{i}_{j}=-\frac{
1}{8}(\theta^{ik}\gamma_{kj}+\delta^{im}\delta_{nj}\gamma_{mk}\theta^{kn})$. The
symplectic structure (\ref{M-twoform}) takes the canonical form
\begin{eqnarray}\label{Dar-M-twoform}
\Omega|_{_{U'}}=dX^i\wedge dY_i\,,
\end{eqnarray}
in this chart. The Darboux transformation (\ref{Dar-MNA}) is linear and clearly does not
impose any restriction on the canonical variables $(X^i,Y_i)$. These variables are not
bounded similar to the positions and momenta $(q_0^i,{p_0}_i)$ of particles in standard
Hamiltonian systems. Note that the Hamiltonian function should be function of
deformation parameters as $H|_{_U'}=H(\theta,\gamma;X,Y)$ in order to guaranty that the
trajectories being the same on ${\mathcal M}$. Substituting (\ref{Dar-M-twoform}) into
the relation (\ref{HVF}) gives
\begin{eqnarray}\label{Dar-MHamilton}
\frac{dX^i}{dt}=\frac{\partial H(\theta,\gamma)}{\partial Y_i},
\hspace{2cm}\frac{dY_i}{dt}=-\frac{\partial H(\theta,\gamma)}{
\partial X^i}.
\end{eqnarray}
Indeed, we have a unique Hamiltonian system which is represented in two different charts
$U$ and $U'$ with noncanonical $(q^i,p_i)$ and canonical variables $(X^i,Y_i)$
respectively.

The Moyal-deformed Hamiltonian system with noncanonical algebra (\ref{moyal-algebra})
is locally quite different from the standard Hamiltonian system with Poisson algebra
(\ref{PA0}) since the trajectories on the Moyal-deformed phase space ${\mathcal M}$ are
different (leading to the modified Hamilton's equations (\ref{MHamilton}) or
equivalently (\ref{Dar-MHamilton})) from the trajectories on the standard phase space
${\mathcal M}_0$. In the low energy limit $\theta^{ij} ,\eta^i_j,\gamma_{ij}\rightarrow0
$, these trajectories coincide. From global point of view, however, the Moyal-deformed
algebra (\ref{moyal-algebra}) is defined on a symplectic manifold with trivial
${\mathcal M}={\mathbb R}^{2n}$ topology similar to the standard Hamiltonian system.
This is because the Moyal-deformed Hamiltonian system does not induce any cutoff for the
system under consideration.

\subsection{Snyder algebra}
In the pervious subsection we considered an example of a deformed Hamiltonian system
that does not induce any cutoff. In this subsection we consider the Snyder-deformed
phase space which induces a UV cutoff for the system under consideration.

Inspired by the seminal work of Snyder on discrete spacetime with noncommutative
coordinates \cite{Snyder}, a locally-deformed noncanonical Poisson algebra is suggested
in non-relativistic limit \cite{Snyderclass}. This algebra also coincides with one that
is arisen from the generalized uncertainty principle in the context of string theory
\cite{GUP-classical}. For the sake of simplicity, we restrict our consideration to the
two-dimensional phase space which is sufficient for our purpose in this paper. The
corresponding Snyder-deformed Poisson algebra is given by
\begin{equation}\label{Snyder-algebra}
\{q,p\}|_{_U}=1+\sigma^2p^2\,,
\end{equation}
where $\sigma$ is the deformation parameter with a dimension of length which expect to
be of the order of Planck length \cite{QGExpriment}. It is easy to show that this
algebra can be generated from the locally-deformed symplectic structure
\begin{equation}\label{Snyder-symplectic}
\Omega|_{_U}=\frac{dq\wedge{dp}}{1+\sigma^2p^2}\,,
\end{equation}
through the definition (\ref{PBD}). The Hamiltonian system is defined by
$({\mathcal M},\Omega,X_{_H})$ with the standard form of the Hamiltonian function as
\begin{equation}\label{Snyder-Hamiltonian}
H|_{_U}=\frac{p^2}{2m}+U(q)\,,
\end{equation}
where $U(q)$ is the potential function. Substituting (\ref{Snyder-symplectic}) into the
relation (\ref{HVF}) gives
\begin{equation}\label{Snyder-VF}
X_{_H}|_{_U}=(1+\sigma^2p^2)\left(\frac{\partial{H}}{\partial{p}
}\frac{\partial}{\partial{q}}-\frac{\partial{H}}{\partial{q}}
\frac{\partial{H}}{\partial{p}}\right)\,,
\end{equation}
and the deformed Hamilton's equation are given by
\begin{eqnarray}\label{Snyder-Hamilton}
\frac{dq}{dt}=(1+\sigma^2p^2)\frac{\partial{H}}{\partial{p}},\hspace{1.5cm}
\frac{dp}{dt}=-(1+\sigma^2p^2)\frac{\partial{H}}{\partial{q}}.
\end{eqnarray}
In the limit of $\sigma\rightarrow0$, all of the above relations reduce to the standard
canonical Hamiltonian system $({\mathcal M}_0,\Omega_0,X_{_{H_0}})$. Up to now all of
the results are the same as the Moyal algebra considered in the pervious subsection.
However, interesting results aries when one considers the total volume (\ref{Tot-Vol})
of the phase space in this setup. The determinant of symplectic structure
(\ref{Snyder-symplectic}) is a function of momentum and potentially could make the total
volume to be finite\footnote{Note that the Snyder algebra (\ref{Snyder-algebra}) imposes
a UV cutoff and not IR cutoff and, therefore, just the momentum part of the phase space
volume will be finite. In a similar manner one could also consider an IR cutoff
which makes the spatial part to be finite. Here, for the sake of simplicity, we suppose
that the spatial volume is restricted to a physical volume in order to get ride of the
IR divergences.}. The total volume (\ref{Tot-Vol}) in this setup is given by
\begin{equation}\label{Snyder-Tot-Vol}
\mbox{Vol}(\Omega)|_{_U}=\int_{V}dq\times\int_{-\infty}^{+\infty}\frac{dp}{1+
\sigma^2{p^2}}=V\times\Big(\frac{\pi}{\sigma}\Big)\,,
\end{equation}
where $V$ is the one-dimensional spatial volume of the system under consideration.
According to our results, there will be a nonlinear Darboux map which transforms the
deformed noncanonical algebra (\ref{Snyder-algebra}) to a canonical one but with bounded
variables. The particular form of the Darboux map (\ref{Darb-trans}) from chart $U$
with noncanonical variables $(q,p)$ to chart $U'$ with canonical variables $(X,Y)$ for
the Snyder algebra (\ref{Snyder-algebra}) is given by
\begin{equation}\label{Snyder-Darb-trans}
(q,p)\rightarrow\Big(X=q,\,Y=\frac{1}{\sigma}\tan^{-1}\sigma{p}\Big)\,,
\end{equation}
which locally transforms the symplectic structure (\ref{Snyder-symplectic}) to the
canonical form
\begin{equation}\label{Snyder-Dar-symplectic}
\Omega|_{_{U'}}=dX\wedge{dY}\,.
\end{equation}
In this example, only a UV cutoff considered and we then concentrate to the momentum
part of the symplectic manifold. The Darboux transformation (\ref{Snyder-Darb-trans})
for the momentum sub-manifold is {\it nonlinear} and also the associated new variable
$Y$ is bounded as $Y\in(-\frac{\pi}{2\sigma},+\frac{\pi}{2\sigma})$ in agreement with
our proposition in the pervious section. The total volume (\ref{Tot-Vol}) is invariant
under the local transformation and calculating it in the new chart $U'$ gives $\mbox{
Vol}(\Omega)|_{_{U'}}=\int_{V}dX\times\int_{-\frac{\pi}{2\sigma}}^{+\frac{\pi}{2\sigma
}}dY=V\times(\pi/\sigma)$ the same result as (\ref{Snyder-Tot-Vol}). Note that $Y$ is
not momentum of a particle but rather a new coordinate on chart $U'\in{\mathcal M}$.
In new chart $U'$, the Hamiltonian function clearly becomes a function of deformation
parameter $\sigma$ as
\begin{equation}\label{Snyder-Dar-Hamiltonian}
H|_{_{U'}}=\frac{\tan^2{\sigma{Y}}}{2m\sigma^2}+U(X)\,.
\end{equation}
The Darboux transformation (\ref{Snyder-Darb-trans}) with new bounded variable $Y$
suggests the compactification of the momentum space to a circle $S^1$. Therefore, the
topology of the phase space on which the Snyder-deformed algebra (\ref{Snyder-algebra})
is defined will be ${\mathcal M}={\mathbb R}\times{S^1}$ in agreement with definition
for the cutoff-regularized Hamiltonian system in the pervious section. In the next
subsection we consider a polymerized phase space with the standard non-deformed canonical
Poisson algebra that is defined on a momentum space with compact $S^1$ topology. We will
see that the UV cutoff naturally arises without any references to a noncanonical Poisson
algebra.

\subsection{Polymerized phase space}
The so-called polymer quantization is investigated in the symmetric sector of loop
quantum gravity \cite{Ashtekar}. At the classical regime, this quantum theory leads to
an effective theory which supports the existence of a minimal length scale known as the
polymer length scale \cite{Corichi1}. Evidently, this classical effective theory can be
obtained through a process known as the polymerization \cite{Corichi2}. The momentum is
compactified to a circle $S^1$ in this setup and there is a maximal momentum
(corresponding to a polymer length) for the system under consideration. The Hamiltonian
system is then defined on symplectic manifold with topology ${\mathcal M}={\mathbb R}
\times{S^1}$. In contrast to the above mentioned models, the Poisson algebra takes the
standard canonical form
\begin{equation}\label{Poly-algebra}
\{q,p\}|_{_U}=1\,,
\end{equation}
in terms of position $q$ and momentum $p$ of particle in this setup. The Poisson algebra
(\ref{Poly-algebra}) can be generated by the symplectic structure
\begin{equation}\label{Poly-symplectic}
\Omega|_{_U}=dq\wedge{dp}\,,
\end{equation}
through the definition (\ref{PBD}). Up to this moment, the system is similar to the
standard Hamiltonian system $({\mathcal M}_0,\Omega_0,X_{_{H_0}})$. The deformation,
however, encoded in the Hamiltonian function that is deformed as \cite{Corichi2}
\begin{equation}\label{Poly-Hamiltonian}
H(\lambda;q,p)|_{_U}=\frac{\sin^2{\lambda{p}}}{2m\lambda^2}+U(q).
\end{equation}
where $\lambda$ is the polymer length scale and $U(q)$ is the potential function. The
momentum is bounded as $p\in[-\frac{\pi}{2\lambda},+\frac{\pi}{2\lambda})$ to have a
single-valued Hamiltonian function. The associated Liouville volume coincides with
symplectic structure (\ref{Poly-symplectic}) and the total volume of the
two-dimensional polymeric phase space will be
\begin{eqnarray}\label{Poly-TV}
\mbox{Vol}(\Omega)|_{_U}=\int_{V}dq\times\int_{-\frac{\pi}{2\lambda}}^{
+\frac{\pi}{2\lambda}}dp=V\times\Big(\frac{\pi}{\lambda}\Big),
\end{eqnarray}
where $V$ is the one-dimensional spatial volume of the system under consideration.
Note that the momentum part of the polymerized phase space is UV-regularized since
it is defined on a momentum space with compact topology. This result is in agreement
with the definition of cutoff-regularized Hamiltonian system in the pervious section.

Substituting (\ref{Poly-symplectic}) into (\ref{HVF}), gives the following solution
for the Hamiltonian vector field
\begin{eqnarray}\label{Poly-VF}
X_{_H}|_{_U}=\frac{\sin{2\lambda{p}}}{2m\lambda}\frac{\partial}{
\partial{q}}-\frac{\partial{U}}{\partial{q}}\frac{\partial}{
\partial{p}}.
\end{eqnarray}
The Hamilton's equations are then modified as
\begin{eqnarray}\label{Poly-Hamilton}
\frac{dq}{dt}=\frac{\sin{2\lambda{p}}}{2m\lambda},\hspace{.5cm}
\frac{dp}{dt}=-\frac{\partial{U}}{\partial{q}}.
\end{eqnarray}
The polymerized Hamiltonian system is globally and locally different from the standard
Hamiltonian system $({\mathcal M}_0,\Omega_0,X_{_{H_0}})$. Note that the Hamilton's
equations (\ref{Poly-Hamilton}) are different from the standard Hamilton's equations
(\ref{Hamilton0}) and the latter can be recovered in the low energy limit of $\lambda
\rightarrow\,0$.

All of the results of this section are summarized in Table \ref{tab:1}. The results
show that from the global point of view there is no difference between standard
Hamiltonian systems and the Hamiltonian system associated to the constant
Moyal-deformed algebra (\ref{moyal-algebra}). Both of them are defined on phase
space with standard ${\mathbb R}^{2n}$ topology and could not exert any cutoff for
the system under consideration. However, these two Hamiltonian systems are locally
quite different. For instance, the expectation values of the physical quantities
(\ref{Expectation-Obs}) are different in this two Hamiltonian systems and also
the trajectories are different such that the standard Hamiltonian system can be
recovered in the low energy limit of the Moyal-deformed algebra. For the two other
cases, Snyder and polymer deformed algebras, the results are more interesting.
These systems locally and globally are different from the standard Hamiltonian
systems. But, they are globally the same such that both of them are defined on
two-dimensional phase spaces with the same ${\mathbb R}\times {S^1}$ topology, where
${\mathbb R}$ is identified with the space of position and $S^1$ with the space of
momentum. Since the topology of the momentum part is compact, the volume of the
momentum space is finite in both cases and there is a natural UV cutoff for the
system under consideration. From the local viewpoint, however, these systems are
quite different. Although the total volume (\ref{Tot-Vol}) leads to the same
expressions (\ref{Snyder-Tot-Vol}) and (\ref{Poly-TV}), the Hamiltonian functions
and symplectic structures take different functional forms even one compares them in
the noncanonical chart or canonical coordinates (see the functional form of
Hamiltonian function in relations (\ref{Snyder-Dar-Hamiltonian}) for Snyder case
and (\ref{Poly-Hamiltonian}) in polymer framework.). Therefore, the trajectories are
different in these two setups and also the expectation values of the physical
quantities are different. In order to show the physical importance of this local
difference, let us consider the expectation value of the kinetic energy $K$ in these
two setups. At the quantum level, in momentum polarization the Hilbert spaces are
${\mathrm H}=L^2(S^1,dp/1+\sigma^2p^2)$ and ${\mathrm H}=L^2(S^1,dp)$ for the Snyder
and polymer cases respectively. Therefore, for the case of the Snyder-deformed
Hamiltonian system with symplectic structure (\ref{Snyder-symplectic}) and Hamiltonian
function (\ref{Snyder-Hamiltonian}), it is diverging as
\begin{equation}\label{Snyder-kinetic}
\langle{K}\,\rangle|_{_U}=\frac{\int_{_{{\mathbb R}
\times{S^1}}}\,\Big(\frac{p^2}{2m}\Big)\Omega_{_{\rm Snyder}}
}{\int_{_{{\mathbb R}\times{S^1}}}\,\Omega_{_{\rm Snyder}}}=
\frac{\sigma}{2m\pi}\int_{-\infty}^{+\infty}\frac{p^2dp}{1+
\sigma^2p^2}\rightarrow\infty\,,
\end{equation}
while for the case of the polymerized phase space with symplectic structure
(\ref{Poly-symplectic}) and Hamiltonian function (\ref{Poly-Hamiltonian}), it is
convergent as
\begin{equation}\label{Polymer-kinetic}
\langle{K}\,\rangle|_{_U}=\frac{\int_{_{{\mathbb R}
\times{S^1}}}\,\Big(\frac{\sin^2\lambda{p}}{2m\lambda^2}\Big)
\Omega_{_{\rm Polymer}}}{\int_{_{{\mathbb R}\times{S^1}}}\,
\Omega_{_{\rm Polymer}}}=\frac{1}{2m\pi\lambda}\int_{-\frac{
\pi}{2\lambda}}^{+\frac{\pi}{2\lambda}}\sin^2\lambda{p}dp=
\frac{1}{4m\lambda^2}\,.
\end{equation}
These simple calculations show the importance of the local deformations from the physical
point of view even if there exist a cutoff. The relation (\ref{Snyder-kinetic}) is
exactly the classical counterpart of the relation (26) of Ref. \cite{Kempf} in which
quantization of the Snyder-deformed Hamiltonian system on the Hilbert space
${\mathrm H}=L^2(S^1,dp/1+\sigma^2p^2)$ is studied. The relation (\ref{Polymer-kinetic})
is also classical counterpart of the relation (42) of Ref. \cite{snyder-poly} where
the quantization of the polymerized phase space is studied (note however that in Ref.
\cite{snyder-poly}, the expectation value is calculated on a noncanonical chart).

\begin{table*}
\caption{\label{tab:1} Three examples of the deformed Hamiltonian
systems are compared from the local and global points of view. The
results show that natural cutoffs arise when the phase space has
compact topology. This result promote cutoffs to be a global feature
of Hamiltonian systems. Note that all of the local results are
considered in physical chart $U$ in which the local coordinates are
positions and momenta of particles.}\vspace{.3cm}
\begin{tabular}{ccccccc}
 Hamiltonian&Poisson&Hamiltonian&Local&Topology&Global&Cutoff-\\
 system&Algebra&function&deformation&&deformation&regularized\\
 \hline Standard&Canonical&Standard&No&Noncompact&No&No\\Moyal&
 Noncanonical&Standard&Yes&Noncompact&No&No\\Snyder&Noncanonical
 &Standard&Yes&Compact&Yes&Yes\\Polymerized&Canonical&Deformed
 &Yes&Compact&Yes&Yes
\end{tabular}
\end{table*}

The results that are obtained in this and pervious sections show that topology of the
momentum part of a two-dimensional phase space will be compactified to a circle $S^1$
in order to have a UV cutoff. What is the generalization of this result for more
natural case of the six-dimensional phase space with three-dimensional momentum space?
Taking only the compactness criterion cannot completely determine topology of the
three-dimensional momentum space. For instance, there is not a clear reason to choose
$S^3$ or $T^3$ topologies. Note, however, that there will be just one parameter of
deformation corresponding to the maximal momentum (UV cutoff). In this respect,
$T^3=S^1\times{S^1}\times{S^1}$ is not an appropriate
choice since it takes into account three deformation parameters correspond to three
maximal momenta for three component of momenta which breaks the rotational invariance.
Therefore, the natural generalization of $S^1$ topology to the three-dimensional case
is $S^3$ topology. The maximally symmetric three-sphere space with $S^3$ topology has
constant curvature which is consistent with the fact that the UV cutoff will be a
universal quantity. Interestingly, this result is also suggested by the doubly special
relativity theories by very different assumptions. The doubly special relativity
theory was suggested by Camelia by means of the deformation of the Lorentz
transformations such that it take into account a minimal observer-independent length
scale (UV cutoff) \cite{DSR}. This suggests that the Lorentz symmetry may be broken
at the very high energy regime . It was later shown that the Lorentz symmetry can be
preserved in this setup \cite{DSR2} and furthermore it is shown that the momentum
space of the underlying system should have de Sitter geometry rather than the standard
Minkowski geometry \cite{Glikman}. Recently, this hypothesis is extended also to the
case of anti-de Sitter space to be a candidate for the momentum space of the doubly
special relativity theories \cite{AdS}. Note that the topologies of de Sitter and
anti-de Sitter spaces are ${\mathbb R}\times{S}^3$ and $S^1\times{\mathbb R}^3$
respectively. In this respect, a maximal momentum arises by a reasonable identification
of ${\mathbb R}$ with the space of energy and $S^3$ with the space of momenta in de
Sitter case and a maximal energy arises by identification of $S^1$ with the space of
energy and ${\mathbb R}^3$ with the space of momenta in the case of anti-de Sitter
momentum space. Moreover, the local coordinates are very important in these setups
which lead to the different doubly special relativity theories \cite{AdS,Glikman}.
These results are exactly in agreement with the results that are obtained in this
paper.

\section{Conclusions and Remarks}
The deformed Hamiltonian systems are usually implemented to take into account natural
ultraviolet (UV) and infrared (IR) quantum gravity cutoffs in the context of
phenomenological quantum gravity models such the generalized uncertainty principle,
polymerized phase space, noncommutative phase spaces and doubly special relativity
theories. These setups are usually deal with local deformations of Hamiltonian systems
leading to a noncanonical Poisson algebra or a deformed Hamiltonian function (modified
dispersion relation). In order to explore how these theories can induce natural cutoffs,
we consider the deformed Hamiltonian system $({\mathcal M},\Omega,X_{_H})$ to be quantum
gravity counterpart of the standard Hamiltonian system $({\mathcal M}_0,\Omega_0,
X_{_{H_)}})$ such that the latter being the low energy limit of the former in the light
of the correspondence principle. Thus, we start from the local deformation of symplectic
structure as (\ref{omega}) in physical chart $U\in{\mathcal M}$ (the chart in which the
local coordinates $(q^i,p_i)$ are the positions and momenta of particles). This
deformation leads to the locally-deformed noncanonical Poisson algebra (\ref{NPA})
which is the starting point of the phenomenological models of quantum gravity. We then
represent the Hamiltonian system $({\mathcal M},\Omega,X_{_H})$ in another local
(Darboux) chart $U'\in{\mathcal M}$ with local coordinates $(X^i,Y_i)$ in which the
Poisson algebra locally takes the canonical form (\ref{Dar-CPA}). This result shows that
while the quantum gravity phenomenological models induce cutoffs by means of noncanonical
Poisson algebras of the form (\ref{NPA}), this claim cannot be treated as the origin of
cutoffs. We then define the cutoffs in a chart-independent manner such that the
cutoff-regularized system should have a finite total phase space volume. We show that if
a noncanonical Poisson algebra (\ref{NPA}), represented in chart $U$ with unbounded
local coordinates $(q^i,p_i)$, makes the total volume (\ref{Tot-Vol}) of the symplectic
manifold ${\mathcal M}$ to be finite, there is a nonlinear Darboux map
(\ref{Darb-trans}) from chart $U$ with coordinates $(q^i,p_i)$ to chart $U'$ with
bounded local coordinates $(X^i,Y_i)$ in terms of which the Poisson algebra takes the
canonical form (\ref{StrucDarb}). But, how the phase space variables can be bounded in
a local chart? The answer is when the phase space have compact topology rather than the
trivial ${\mathbb R}^{2n}$ topology. In this respect, we find that the deformed
Hamiltonian system $({\mathcal M},\Omega,X_{_H})$ should be not only locally but also
globally different from the standard Hamiltonian system
$({\mathcal M}_0,\Omega_0,X_{_{H_0}})$. While the standard Hamiltonian systems are
defined on trivial ${\mathbb R}^{2n}$ topology, their quantum gravity counterpart
$({\mathcal M},\Omega,X_{_H})$ will be defined only on a compact symplectic manifold.
Thus, the symplectic structure is closed but not exact 2-form on ${\mathcal M}$ while
its closed and exact on ${\mathcal M}_0$. This result shows that the cutoffs are
related to the global properties of the symplectic manifolds. Nevertheless, it is
important to note that while the local deformations play no role to induce cutoffs,
they are important from the physical point of view. In fact, the expectation values of
the physical quantities are directly determined by the local form of the deformation of
the Hamiltonian systems and also the trajectories on the phase space are directly
affected by these local deformations. In order to justify our results, we considered
three examples of the deformed Hamiltonian systems:

\begin{itemize}
  \item {\it Moyal-deformed Hamiltonian system:} The Poisson algebra
  (\ref{moyal-algebra}) is noncanonical in this setup and this system is locally
  different from the standard Hamiltonian system. These two system, however, are
  globally the same such that both of them are defined on trivial ${\mathbb{R}}^{2n}$
  topology and therefore cannot induce any cutoffs for the system under consideration.

  \item {\it Snyder-deformed Hamiltonian system:} The Poisson algebra
  (\ref{Snyder-algebra}) is noncanonical in this setup similar to the Moyal-deformed
  case. This system is however locally and also globally different from the
  Moyal-deformed case. It is defined on phase space with compact $S^1$ momentum space
  and, therefore, induces a UV cutoff.

  \item {\it Polymerized Hamiltonian system:} The Poisson algebra (\ref{Poly-algebra})
  is canonical but the momentum space is $S^1$. The UV cutoff naturally arises in
  this setup without any reference to a noncanonical Poisson algebra. The polymerized
  phase space and the Snyder-deformed system are globally indistinguishable but
  locally are different.
\end{itemize}
All of the local properties in the above examples are considered in physical chart $U$
in which the local coordinates are positions and momenta of particles. These results
are summarized in table \ref{tab:1}. We also calculate the average of kinetic energy
in both of the Snyder and polymerized phase spaces which are given by the relations
(\ref{Snyder-kinetic}) and (\ref{Polymer-kinetic}) respectively. While both of these
Hamiltonian systems are UV-regularized, the average of the kinetic energy diverges for
the Snyder case and converges in the polymerized phase space. This result show the
importance of local deformations for the cutoff-regularized Hamiltonian systems.

Now we compare this framework with general relativity: Gravity is universal in the
sense that it couples to everything and is described by curvature of spacetime manifold
through the equivalence principle. This means that there is no flat spacetime in essence
and therefore gravity can be interpreted as a local property of Riemannian manifold. In
a similar manner, natural cutoffs are universal in the sense that all physical systems
have uniquely affected by them. For instance, a minimal length scale (corresponding to
UV cutoff) cannot be Lorentz contracted since it is the same for any observer
\cite{DSR,DSR2,spallucci}. We have shown that natural cutoffs can be realized from
compact topology of symplectic manifolds. Thus, in comparison with gravity, we can
say that natural cutoffs are global properties of the symplectic manifolds. While the
classical theory of gravity, {\it i.e.} general relativity, is a local theory, it seems
that at the quantum level global effects are important.\\

{\bf Acknowledgement}\\
We would like to thank the referees for very insightful comments which greatly improved
the quality of the paper.

\end{document}